\documentclass[nohyperref]{article}

\usepackage{microtype}
\usepackage{graphicx}
\usepackage{subfigure}
\usepackage{booktabs} 

\usepackage{hyperref}

\usepackage[]{algorithm2e}
\usepackage{bibentry}
\usepackage{array}
\usepackage{verbatim}
\usepackage{enumitem}
\usepackage{amssymb}
\usepackage{mathtools}

\usepackage[accepted]{icml2022}

\usepackage{amsmath}
\usepackage{amssymb}
\usepackage{mathtools}
\usepackage{amsthm}
\SetKwInput{kwInput}{\textbf{Input}}
\SetKwInput{kwInit}{\textbf{Initialization}}
\usepackage[capitalize,noabbrev]{cleveref}

\theoremstyle{plain}

\theoremstyle{definition}

\theoremstyle{remark}

\newcommand{\dbest}[0]{\mathcal{D}_{\text{Best}}}
\newcommand{\Tau}{\mathcal{T}}
\DeclarePairedDelimiter\ceil{\lceil}{\rceil}

\usepackage[textsize=tiny]{todonotes}

\icmltitlerunning{Biological Sequence Design with GFlowNets}

\begin{document}

\twocolumn[
\icmltitle{Biological Sequence Design with GFlowNets}




\begin{icmlauthorlist}
\icmlauthor{Moksh Jain}{mila,udem}
\icmlauthor{Emmanuel Bengio}{mila,mcgill}
\icmlauthor{Alex-Hernandez Garcia}{mila,udem}
\icmlauthor{Jarrid Rector-Brooks}{mila,udem}
\icmlauthor{Bonaventure F. P. Dossou}{mila,udem,jacobs}
\icmlauthor{Chanakya Ekbote}{mila}
\icmlauthor{Jie Fu}{mila,udem}
\icmlauthor{Tianyu Zhang}{mila,udem}
\icmlauthor{Michael Kilgour}{nyu}
\icmlauthor{Dinghuai Zhang}{mila,udem}
\icmlauthor{Lena Simine}{mcgill}
\icmlauthor{Payel Das}{ibm}
\icmlauthor{Yoshua Bengio}{mila,udem,cifar}
\end{icmlauthorlist}

\icmlaffiliation{mila}{Mila}
\icmlaffiliation{mcgill}{McGill University}
\icmlaffiliation{jacobs}{Jacobs University Bremen}
\icmlaffiliation{nyu}{New York University}
\icmlaffiliation{udem}{Universit\'e de Montr\'eal}
\icmlaffiliation{ibm}{IBM}
\icmlaffiliation{cifar}{CIFAR Fellow and AI Chair}
\icmlcorrespondingauthor{Moksh Jain}{mokshjn00@gmail.com}
\icmlkeywords{Machine Learning, ICML}

\vskip 0.3in
]



\printAffiliationsAndNotice{}  

\begin{abstract}
Design of \emph{de novo} biological sequences with desired properties, like protein and DNA sequences, often involves an active loop with several rounds of molecule ideation and expensive wet-lab evaluations. These experiments can consist of multiple stages, with increasing levels of precision and cost of evaluation, where candidates are filtered. This makes the diversity of proposed candidates a key consideration in the ideation phase. In this work, we propose an active learning algorithm leveraging epistemic uncertainty estimation and the recently proposed GFlowNets as a generator of diverse candidate solutions, with the objective to obtain a diverse batch of useful (as defined by some utility function, for example, the predicted anti-microbial activity of a peptide) and informative candidates after each round. We also propose a scheme to incorporate existing labeled datasets of candidates, in addition to a reward function, to speed up learning in GFlowNets. We present empirical results on several biological sequence design tasks, and we find that our method generates more diverse and novel batches with high scoring candidates compared to existing approaches. 
\end{abstract}

\section{Introduction}
\label{sec:intro}
Biological sequences like proteins and DNA have a broad range of applications to several impactful problems ranging from medicine to material design. 
For instance, design of novel anti-microbial peptides (AMPs; short sequences of amino-acids) is crucial, and identified as the first target to tackle the growing public health risks posed by increasing anti-microbial resistance \citep[AMR; ][]{2022amr}. 
\setcounter{footnote}{1}
This is particularly alarming according to a recent report\footnote{\smaller\url{https://www.who.int/news-room/fact-sheets/detail/antibiotic-resistance}} by the World Health Organization, which predicts millions of human lives lost per year (with the potential breakdown of healthcare systems and many more indirect deaths), unless methods to efficiently control (and possibly stop) the fast-growing AMR are found. 

\begin{figure}[t]
    \centering
    \includegraphics[width=1\linewidth]{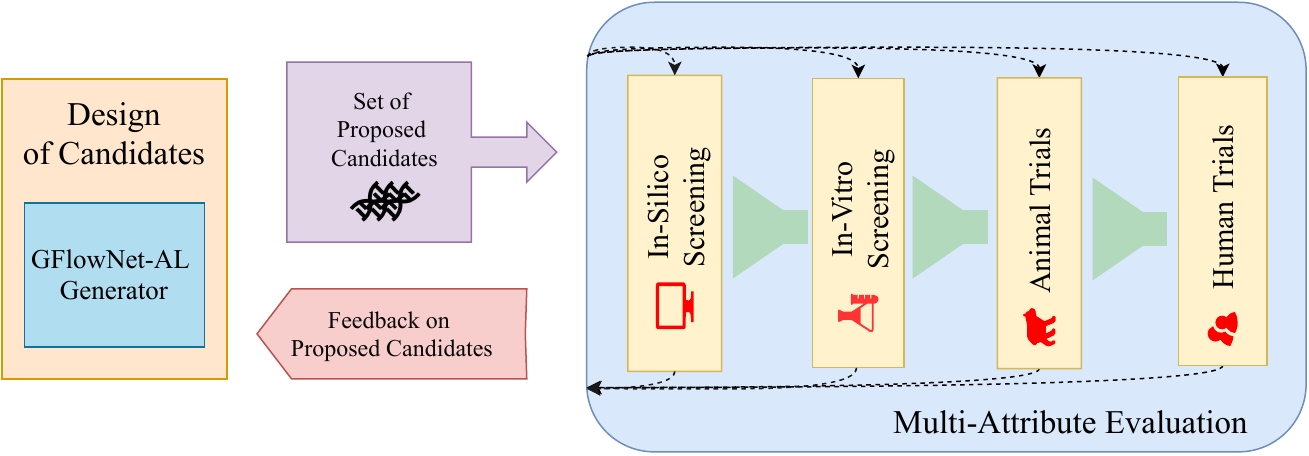}
    \caption{Illustration of a typical drug discovery pipeline. In each round, a set of candidates is proposed which are evaluated under various stages of evaluation, each measuring different properties of the candidates with varying levels of precision.
    The design procedure is then updated using the feedback received from the evaluation phase before the next round begins. Because the early screening phases are imperfect, and the ideal ``usefulness" of the candidate can be ill-defined, it is important to generate for these phases a {\em diverse} set of candidates (rather than many similar candidates who could all fail in the downstream phases).}
    \label{fig:ddpipeline}
\end{figure}

Considering the diverse nature of the biological targets, modes of attack, structures, as well as the evolving nature of such problems, diversity becomes a key consideration in the design of these sequences \cite{mullis2019diversity}.
Another reason for the importance of being able to propose a {\em diverse} set of good candidates is that cheap screening methods (like in-silico simulations or in-vitro experiments) may not reflect well future outcomes in animals and humans, as illustrated in Figure~\ref{fig:ddpipeline}. To maximize the chances that at least one of the candidates will work in the end, it is important for these candidates to cover as much as possible the modes of a \emph{goodness} function that estimates future success. 
The design of new biological sequences involves searching over combinatorially large discrete search spaces on the order of $\mathcal{O}(10^{60})$ candidates. Machine learning methods that can exploit the combinatorial structure in these spaces (e.g., due to laws of physics and chemistry) have the potential to speed up the design process for such biological sequences \cite{bayes4DD, bayesDD2, das2021accelerated}. 

The development process of such biological sequences, for a particular application, involves several rounds of a candidate ideation phase (possibly starting with a random library) followed by an evaluation phase, as shown in Figure~\ref{fig:ddpipeline}. The evaluation consists of several stages ranging from numerical simulations to expensive wet-lab experiments, possibly culminating in clinical trials. These stages filter candidates with progressively higher fidelity oracles that measure different aspects of the \textit{usefulness} of a candidate. For example, the typical evaluation for an antibiotic drug after ideation would comprise of: 
(1) in-silico screening using approximate models to estimate anti-microbial activity of $\mathcal{O}(10^6)$ candidates 
(2) in-vitro experiments to measure single-cell effectiveness against a target bacterium species of $\mathcal{O}(10^3)$ candidates (3) trials in small mammals like mice with $\mathcal{O}(10)$ $\,$candidates 
(4) randomized human trials with $\mathcal{O}(1)$ candidates. 
These oracles are often imperfect and do not evaluate all the required properties of a candidate.

The biological repertoire of DNA, RNA and protein sequences is extremely diverse, to support the diversity of structure, function and modes of action exploited by living organisms, where the same high-level function can be potentially executed in more than one possible manner~\citep{mullis2019diversity}.
Moreover, the ultimate success of candidate drugs also depends on satisfying multiple often conflicting desiderata, not all of which likely can be precisely estimated in-silico. This fact, combined with
the overall effect of the above aggressive filtering and use of potentially imperfect oracles, needs to be addressed in the design phase through the \emph{diversity} of the generated candidates. 
Diverse candidates capturing the \emph{modes} of the imperfect oracle improve the likelihood of discovering a candidate that can satisfy all (or many) evaluation criteria, because failure in downstream stages is likely to affect nearby candidates (from the same mode of the oracle function), while different modes are likely to correspond to qualitatively different properties.

This setup of iteratively proposing a batch of candidates and learning from the feedback provided by an oracle on that batch fits into the framework of active learning \cite{aggarwal2014active}. 
Bayesian Optimization is a common approach for such problems \cite{rasmussen2005gaussian,garnett_bayesoptbook_2022}.
It relies on a Bayesian surrogate model of the usefulness function of interest (e.g., the degree of binding of a candidate drug to a target protein), with an output variable $Y$ that we can think of as a reward for a candidate $X$.  An acquisition function ${\cal F}$ is defined on this surrogate model and a pool of candidates will be screened to search for candidates $x$ with a high value of ${\cal F}(x)$. That acquisition function combines the expected reward function $\mu$ (e.g., $\mu(x)$ can be the probability of obtaining a successful candidate) as well as an estimator of epistemic uncertainty $\sigma(x)$ around $\mu(x)$, to favour candidates likely to bring new information to the learner. 
There are many possible candidate selection procedures, from random sampling to genetic algorithms evolving a population of novel candidates \cite{bayes4DD,belanger2019biological, moss2020bayesian, swersky20amortized, bayesDD2}. An alternative is to use Reinforcement Learning (RL) to maximize the value of a surrogate model of the oracle \cite{angermueller2019model}. 
RL methods are designed to search for a single candidate that maximizes the oracle, which can result in poor diversity and can cause candidate generation to get \emph{stuck} in a single mode \cite{bengio2021flow} of the expected reward function. Additionally, as the final goal is to find \emph{novel} designs that are different from the ones that are already known, the generative model must be able to capture the tail ends of the data distribution.

In settings where diversity is important, another interesting way to generate candidates is to use a generative policy that can sample candidates proportionally to a reward function (for instance, the acquisition function over a surrogate model) and can be sampled i.i.d to obtain a set of candidates that covers well the modes of the reward function. 
A sample covering the modes approximately but naturally satisfies the ideal criterion of high scoring and diverse candidates. 
GFlowNets \cite{bengio2021flow} provide a way to learn such a stochastic policy and, unlike Markov chain Monte Carlo (MCMC) methods (which also have this ability), amortize the cost of each new i.i.d. sample (which may require a lengthy chain, with MCMC methods) into the cost of training the generative model~\citep{Zhang2022GenerativeFN}. 
As such, this paper is motivated by the observation that GFlowNets are appealing in the above Bayesian optimization context, compared with existing RL and MCMC approaches in domains such as small molecule synthesis. 

In this work, we present an active learning algorithm based on a GFlowNet generator for the task of biological sequence design. 
In addition to this, we propose improvements to the GFlowNet training procedure to improve performance in active learning settings. 
We apply our proposed approach on a broad variety of biological sequence design tasks. 
The key contributions of this work are summarized below:
\begin{itemize}[itemsep=0pt]
    \item An active learning algorithm with GFlowNet as the generator for designing novel biological sequences.
    \item Investigating the effect of off-policy updates from a static dataset to speed up training of GFlowNets.
    \item Incorporating the epistemic uncertainty in the predicted expected reward to improve exploration in GFlowNets.
    \item Validating the proposed algorithm on three protein and DNA design tasks.
\end{itemize}

\section{Problem Setup}
\label{sec:problem}
We consider the problem of searching over a space of discrete objects $\mathcal{X}$ to find objects $x \in \mathcal{X}$ that maximize a given usefulness measure (oracle) $f: \mathcal{X} \mapsto \mathbb{R}^{+}$. 
We consider the setting where this oracle can only be queried $N$ times in fixed batches of size $b$. This constitutes $N$ rounds of evaluation available to the active learning algorithm. The algorithm also has access to an initial dataset $D_0=\{(x^0_1, y^0_1), \dots, (x^0_n, y^0_n)\}$, where $y^0_i=f(x^0_i)$ from evaluations by the oracle.

The algorithm has to propose a new batch of candidates \mbox{$\mathcal{B}_i = \{x_1^i, \dots, x_b^i\}$}, given the current dataset $\mathcal{D}_i$, in each round $i\in \{1, \dots, N\}$. This batch is then evaluated on the oracle to obtain the corresponding scores for the candidates $y_j^i=f(x_j^i)$. The current dataset $\mathcal{D}_i$ is then augmented with the tuples of the proposed candidates and their scores to generate the dataset for the next round, $\mathcal{D}_{i+1} = \mathcal{D}_i \cup \{(x_1^i, y_1^i), \dots, (x_b^i, y_b^i)\}$.

This problem setup is similar to the standard black-box optimization problem \cite{audet2017derivative} with one difference: the initial dataset $D_0$  is available as a starting point, which is actually a common occurrence in practice, i.e., a historical dataset. This setup can also be viewed as an extension of the Offline Model Based Optimization \cite{trabucco2021conservative,trabucco2021designbench} paradigm to multiple rounds instead of a single round. 

\paragraph{Desiderata for Proposed Candidates}
As discussed in Section~\ref{sec:intro}, searching for a single candidate maximizing the oracle can be problematic in the typical scenario where the available (cheap, front-line) oracle is imperfect. Instead, we are interested in looking for a diverse set of $K$ top candidates generated by the algorithm, $\dbest = \text{TopK}(\mathcal{D}_K \setminus \mathcal{D}_0)$. We outline the key characteristics that define the set of \emph{ideal} candidates. 

\begin{itemize}
    \item \textbf{Performance/Usefulness Score}: The base criteria is for the set to include high scoring candidates, which can be quantified with
    \begin{equation}
    \text{Mean}(\mathcal{D})=\frac{\sum_{(x_i,y_i) \in \mathcal{D}} y_i}{|\mathcal{D}|}
    \end{equation}    
    \item \textbf{Diversity}: In addition to being high scoring, we would like the candidates to capture the modes of the oracle. One way to measure this is 
        
    \begin{equation}
        {\small\text{Diversity}}(\mathcal{D}) \hspace*{-1mm}=\hspace*{-1mm} \frac{\sum\limits_{(x_i,y_i)\in \mathcal{D}}\sum\limits_{(x_j,y_j)\in \mathcal{D}\setminus \{(x_i,y_i)\}}\hspace*{-1mm} d(x_i, x_j)}{|\mathcal{D}|(|\mathcal{D}| - 1) }
    \end{equation} 
    where $d$ is a distance measure defined over $\mathcal{X}$. 
    \item \textbf{Novelty}: Since we start with an initial dataset $\mathcal{D}_0$, the proposed candidates should also be different from the candidates that are already known. We measure this \emph{novelty} in the proposed candidates as follows:
    \begin{equation}
    \text{Novelty}(\mathcal{D}) = \frac{\sum_{(x_i,y_i) \in \mathcal{D}} \min_{s_j \in \mathcal{\mathcal{D}}_0} d(x_i, s_j)}{|\mathcal{D}| }
    \end{equation} 
\end{itemize}

All three metrics are applied on the TopK scoring candidates, i.e., for $\mathcal{D}=\dbest$. It is important to note that either of these metrics considered \emph{alone} can paint a misleading picture. For instance, a method can generate diverse candidates, but these candidates might be low scoring and similar to the known candidates. Thus, a method should be evaluated holistically, considering \emph{all} the three metrics.

\section{GFlowNets For Sequence Design}
\label{sec:gflownets}
\begin{figure}[htbp]
    \centering
    \includegraphics[width=\linewidth]{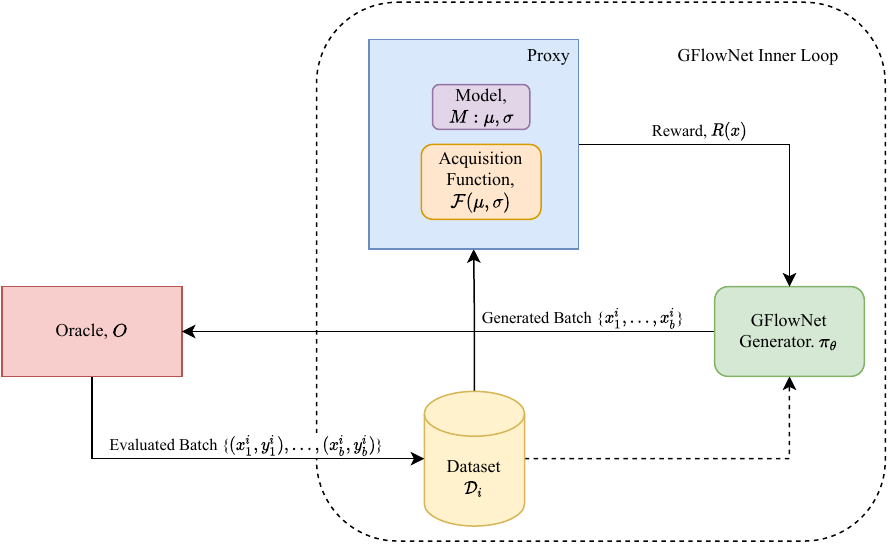}
    \caption{GFlowNet-AL: Our proposed approach for sequence design with GFlowNets consists of three main components: (1) GFlowNet Generator $\pi_\theta$ (green box), which generates diverse candidates with probability proportional to $R(x)$, which is defined by the proxy, (2) Proxy (blue) which consists of a model $M$ that can output a mean prediction $\mu$ and uncertainty estimate $\sigma$ around $\mu$, along with an acquisition function $\mathcal{F}$, which combines the mean and uncertainty predicted by the model, and (3) Dataset $\mathcal{D}_i$ (yellow) which stores all the available candidates up to round $i$. In each round, the model $M$ is first trained on $\mathcal{D}_i$. The generative policy is then trained with reward function $R=\mathcal{F}(M.\mu, M.\sigma)$ and data $\mathcal{D}_i$. A new batch of candidates $\mathcal{B}_i$ is then sampled from $\pi_\theta$, evaluated with the Oracle $\mathcal{O}$ (red) and then added to $\mathcal{D}_i$ to obtain $\mathcal{D}_i$. This process repeats for $N$ rounds of active learning. 
    }
    \label{fig:al_setup}
\end{figure}

GFlowNets \cite{bengio2021flow, bengio2021gflownet} tackle the problem of learning a stochastic policy $\pi$ that can sequentially construct discrete objects $x\in\mathcal{X}$ with probability $\pi(x)$ using a non-negative reward function $R:\mathcal{X}\mapsto \mathbb{R}^{+}$ defined on the space $\mathcal{X}$, such that $\pi(x) \propto R(x)$. This property makes GFlowNets well-positioned to be used as a generator of diverse candidates in an active learning setting. In this section, we present our proposed active learning algorithm based on GFlowNets \cite{bengio2021flow}. We only present the relevant key results, and refer the reader to \citet{bengio2021gflownet} for a thorough mathematical treatment of GFlowNets.   Figure~\ref{fig:al_setup} provides an overview of our proposed approach and Algorithm~\ref{algo:multi_round} describes the details of the approach. 

\begin{algorithm}[h]
\SetAlgoLined
\kwInput{\;\\
$O$: Oracle  to evaluate candidates $x$ and return labels $Y$\;\\
$D_0=\{(x_i, y_i)\}$: Initial dataset with $y_i=O(x_i)$\;\\ 
$M$: Trainable learner providing functions $M.\mu$ and $M.\sigma$, with $\mu(x)$ estimating $E[Y|x]$ and $\sigma(x)$ estimating the epistemic uncertainty around $\mu(x)$\;\\
$\pi_\theta$: Generative policy trainable from a reward function $R$ and from which candidates $x$ can be sampled\;\\ 
$\cal F(\mu,\sigma)$: Acquisition function taking $M.\mu$ and $M.\sigma$ functions and returning a reward function $R$ for training $\pi_\theta$\; \\
$K$: Number of top-scoring candidates to keep for $TopK$ evaluation\; \\
$b$: Size of candidate batch to be generated\; \\
$N$: Number of active learning rounds (outer loop iterations)\;}
\KwResult{$TopK(D_N)$ elements $(x,y) \in D_n$ with highest values of $y$}
\kwInit{ $M$,$\pi_\theta$}
\For{$i=1$ to $N$}{\;\\
$\bullet$ Fit $M$ on dataset $D_{i-1}$\;\\
$\bullet$ Train $\pi_\theta$ with GFlowNet Inner Loop (Algorithm~\ref{algo:inner_loop}) using reward function $R={\cal F}(M.\mu,M.\sigma)$\;\\
$\bullet$ Sample query batch $B = \{x_1, \dots, x_b\}$ with $x_i \sim \pi_\theta$\;\\
$\bullet$ Evaluate batch $B$ with $O$: \mbox{$\hat{D_i}=\{(x_1, O(x_1)), \dots, (x_b, O(x_b))\}$}\;\\
$\bullet$ Update dataset $D_i=\hat{D_i} \cup D_{i-1}$\;
}
 \caption{Multi-Round Active Learning}
\label{algo:multi_round}
\end{algorithm}

\subsection{Background}
\paragraph{Preliminaries}
We assume the space $\mathcal{X}$ is \emph{compositional}, that is, object $x\in\mathcal{X}$ can be constructed using a sequence of actions taken from a set $\mathcal{A}$. After each step, we may have a partially constructed object, which we call a state $s \in \mathcal{S}$. For example, \citet{bengio2021flow} use a GFlowNet to sequentially construct a molecule by inserting an atom or a molecule fragment in a partially constructed molecule represented by a graph. In the auto-regressive case of sequence generation, the actions could just be to append a token to a partially constructed sequence. A special action indicates that the object is complete, i.e., $s = x \in {\mathcal X}$. Each transition $s{\rightarrow}s' \in \mathcal{E}$ from state $s$ to state $s'$ corresponds to an edge in a graph $G= (\mathcal{S}, \mathcal{E})$ with the set of nodes $\mathcal{S}$ and the set of edges $\mathcal{E}$. We require the graph to be directed and acyclic, meaning that actions are constructive and cannot be undone. An object $x \in \mathcal{X}$ is constructed by starting from an initial empty state $s_0$ and applying actions sequentially, and all complete trajectories must end in a special final state $s_f$.
The fully constructed objects in $\mathcal{X}\subset\mathcal{S}$ are  \emph{terminating states}.
The construction of an object $x$ can thus be defined as a trajectory of states $\tau = (s_0 {\rightarrow} s_1 {\rightarrow} \dots{\rightarrow} x {\rightarrow} s_f)$, and we can define $\Tau$ as the set of all trajectories. $\text{Parent}(s) = \{s': s'{\rightarrow} s \in \mathcal{E}\}$ denotes the parents for node $s$ and $\text{Child}(s) = \{s': s{\rightarrow} s' \in \mathcal{E}\}$ denotes the children of node $s$ in $G$. 

\paragraph{Flows}
\citet{bengio2021gflownet} define a \emph{trajectory flow} $F:\Tau \mapsto \mathbb{R}^+$. This trajectory flow $F(\tau)$ can be interpreted as the probability mass associated with trajectory $\tau$. The \emph{edge flow} can then be defined as $F(s{\rightarrow} s') = \sum_{s{\rightarrow} s' \in \tau}F(\tau)$, and \emph{state flow} can be defined as $F(s) = \sum_{s \in \tau}F(\tau)$. The flow associated with the final step (transition) in the trajectory $F(x{\rightarrow}  s_f)$ is called the terminal flow and the objective of training a GFlowNet is to make it approximately match a given reward function $R(x)$ on every possible $x$. 

The trajectory flow $F$ is a measure over complete trajectories $\tau \in \Tau$ and it induces a corresponding probability measure 
\begin{equation}
    P(\tau) = \frac{F(\tau)}{\sum_{\tau \in \Tau}F(\tau)} = \frac{F(\tau)}{Z},
\end{equation}
where $Z$ denotes the total flow, and corresponds to the partition function of the the measure $F$. The forward transition probabilities $P_F$ for each step of a trajectory can then be defined as
\begin{equation}
    P_F(s|s') = \frac{F(s{\rightarrow} s')}{F(s)}.
\end{equation}
We can also define the probability $P_F(s)$ of visiting a terminal state $s$  as 
\begin{equation}
P_F(s) = \frac{\sum_{\tau \in \Tau: s\in\tau}F(\tau)}{Z}.
\end{equation}

\paragraph{Flow Matching Criterion}
A \emph{consistent flow} satisfies the \emph{flow consistency equation} $\forall s \in \mathcal{S}$ defined as follows: 
\begin{equation}
    \sum_{s'\in \text{Parent}(s)} F(s'{\rightarrow} s) = \sum_{s'' \in \text{Child}(s)} F(s{\rightarrow} s'').
\end{equation}

It has been shown \citep{bengio2021flow} that for a consistent flow $F$ with the terminal flow set as the reward, i.e., $F(x{\rightarrow} s_f) = R(x)$, a policy $\pi$ defined by the forward transition probability $\pi(s'|s) = P_F(s'|s)$ samples object $x$ with probability proportional to $R(x)$ 
\begin{equation}
\pi(x) = \frac{R(x)}{Z}.
\end{equation}

\paragraph{Learning GFlowNets}
GFlowNets learn to approximate an \emph{edge flow} $F_\theta:\mathcal{E} \mapsto \mathbb{R}^+$ defined over $G$, such that the terminal flow is equal to the reward $R(x)$ and the flow is \emph{consistent}. This is achieved by defining a loss function whose global minimum gives rise to the consistency condition. 
This was first formulated \citep{bengio2021flow} via a temporal difference-like \citep{sutton2018reinforcement} learning objective, called \emph{flow-matching}:

\begin{equation}
    \mathcal{L}_{FM}(s; \theta) = \left(\log \frac{\sum_{s'\in \text{Parent}(s)} F_\theta(s'{\rightarrow} s)}{\sum_{s'' \in \text{Child}(s)}F_\theta(s{\rightarrow} s'')}\right)^2.
    \label{eq:fm_objective}
\end{equation}

\citet{bengio2021flow} show that given trajectories $\tau_i$ sampled from an exploratory training policy $\tilde{\pi}$ with full support, an edge flow learned by minimizing Equation~\ref{eq:fm_objective} is consistent.
At this point, the forward transition probability defined by this flow $P_{F_\theta}(s'|s) = \frac{F_\theta(s{\rightarrow} s')}{\sum_{s'' \in \text{Child}}F_\theta(s{\rightarrow} s'')}$ would sample objects $x$ with a probability $P_F(x)$ proportionally to their reward $R(x)$.

In practice, the trajectories for training GFlowNets are sampled from an exploratory policy that is a mixture between the GFlowNet sampler $P_{F_\theta}$
and a uniform choice of action among those allowed in each state:
\begin{equation}
\bar{\pi}_\theta = (1-\delta)P_{F_\theta} + \delta \cdot \text{Uniform}.
\end{equation}

This uniform policy introduces exploration preventing the training from getting stuck in one or a few modes. This is analogous to $\epsilon$-greedy exploration in reinforcement learning.

\paragraph{Trajectory Balance} \citet{malkin2022trajectory} present an alternative objective defined over trajectories with faster credit assignment for learning GFlowNets, called \emph{trajectory balance}, defined as follows:

\begin{equation}
    \mathcal{L}_{TB} (\tau;\theta) = \left(\log \frac{Z_\theta \prod_{s{\rightarrow} s' \in \tau}P_{F_\theta}(s'|s)}{R(x)}\right)^2,
    \label{eq:trajbal_objective}
\end{equation}
where $\log Z_\theta$ is also a learnable free parameter. This objective can improve learning speed due to more efficient credit assignment, as well as robustness to long trajectories and large vocabularies. Equation~\ref{eq:trajbal_objective} is the training objective we have used in this paper. 

\paragraph{Remarks}
When generating sequences in an auto-regressive fashion (appending one token at a time), as in this paper, the mapping from trajectories to states becomes \emph{bijective}, as there is only one path to reach a particular state $s$. The directed graph $G$ then corresponds to a directed tree. Under these conditions, the flow-matching objective is equivalent to discrete-action Soft Q-Learning \cite{haarnoja2017reinforcement,buesing2019approximate} with a temperature parameter $\alpha=1$, a uniform $q_{\mathbf{a}'}$, and $\gamma=1$, which obtains $\pi(x)\propto R(x)$. While the trajectory balance objective in \eqref{eq:trajbal_objective} asymptotically reaches the same solution, our results \citep[and that of ][]{malkin2022trajectory} suggest it does so faster.

\begin{algorithm}
\SetAlgoLined
\kwInput{\\
$D=\{x_i, y_i\}, i=1,\dots,N$: Dataset of candidates $x_i$ with known oracle scores $y_i$ \;\\ 
$R(\cdot)$: Reward function\; \\
$\gamma$: Proportion of offline data to use in training\; \\
$m$: GFlowNet training minibatch size\; \\
$T$: number of minibatch updates to complete training\;\\
$\delta$: mixing coefficient for uniform actions in training policy}
\KwResult{$\pi_\theta = P_{F_\theta}$: learned policy with $\pi_\theta(x)\propto R(x)$}
\kwInit{
$F_\theta$: parameterized edge flow (neural net)\;}
\For{$i=1$ to $T$}{\;\\
$\bullet$ Sample $m'=\ceil{m(1-\gamma)}$ trajectories from policy $\tilde{\pi} = (1-\delta)P_{F_\theta} + \delta\, \text{Uniform}$\; \\
$\bullet$ Sample $m-m'\;$ trajectories from dataset $\mathcal{D}$\;\\
$\bullet$ Combine both sets of trajectories to form overall minibatch\;\\
$\bullet$ Compute reward $R(x)$ on terminal states $x$ from each trajectory in the minibatch\; \\
$\bullet$ Update parameters $\theta$ with a stochastic gradient descent step wrt the objective in Eq.~\ref{eq:fm_objective} or Eq.~\ref{eq:trajbal_objective} for all trajectories in the minibatch.
}
 \caption{GFlowNet Inner Loop (with training data)}
\label{algo:inner_loop}
\end{algorithm}

\subsection{Leveraging Data during Training}
In our active learning setting, the reward function for the GFlowNet is obtained by training a model from a dataset $\mathcal{D} = \{(x,y)\}$ of labeled sequences with input object $x$ and observed oracle reward $y$ and we would like to make sure that the GFlowNet samples correctly in the vicinity of these $x$'s (especially those for which $y$ is larger). 
We can observe that the flow-matching objective (Equation~\ref{eq:fm_objective}) and  the trajectory balance objective (Equation~\ref{eq:trajbal_objective}) are \emph{off-policy} and \emph{offline}. This allows us to use trajectories sampled from other policies than $\pi$ during training, so long as the overall distribution of training trajectories $\tilde{\pi}$ has full support. These trajectories can be constructed from the $x$'s in a given dataset by sampling for each of them a sequence of ancestors starting from terminal state $x$ and sampling a parent according to the backward transition probability. In the auto-regressive case studied here, there is only one possible parent for each state $s$, so we immediately recover the unique trajectory leading to $x$ from $s_0$. This provides a set of offline trajectories. 

Inspired by work in RL combining on-policy and off-policy updates \cite{nachum2017bridging, guo2021text}, we propose incorporating trajectories from the available dataset in the training of GFlowNets. At each training step we can augment the trajectories sampled from the current forward transition policy with trajectories constructed from examples in the dataset. Let $\gamma \in [0, 1)$ denote the proportion of offline trajectories in the GFlowNet training batch. As we vary $\gamma$ from $0$ to $1$, we move from an online setting, originally presented in \cite{bengio2021flow}, to an offline setting where we learn exclusively from the dataset. Relying exclusively on trajectories from a dataset, however, can lead to sub-optimal solutions since the dataset is unlikely to cover $\mathcal{X}$.
Algorithm~\ref{algo:inner_loop} describes the proposed training procedure for GFlowNets which incorporates offline trajectories.

We hypothesize and verify experimentally in Section~\ref{sec:mixingresults}, that mixing an empirical distribution in the form of offline trajectories can provide the following potential benefits in the context of active learning:
    {\em (1) improved learning speed}: it can improve the speed of convergence since we make sure the GFlowNet approximation is good in the vicinity of the selected interesting examples from the dataset $\mathcal{D}$
    {\em (2) lower bound on the exploration domain}: it guarantees exploration around the examples in $\mathcal{D}$. 

\subsection{Incorporating Epistemic Uncertainty}

Another consequence of a reward function that is learned from a finite dataset $\mathcal{D} = \{(x,y)\}$ is that there will be increasing uncertainty in the model's predictions as we move away from its training $x$'s. In the context of active learning, this uncertainty can be a strong signal to guide exploration in novel parts of the space and has been traditionally used in Bayesian optimization \cite{angermueller2019model, swersky20amortized,jain2021deup}. \citet{bengio2021gflownet} hypothesize that using information about the uncertainty of the reward function can also lead to more efficient exploration in GFlowNets. We study this hypothesis, by incorporating the model uncertainty of the reward function for training GFlowNets. 

This requires two key ingredients: (a) the reward function should be a model that can provide an uncertainty estimate on its output, and (b) an acquisition function that can combine the prediction of the reward function with its uncertainty estimates to provide a scalar score. There has been significant work in developing methods that can estimate the uncertainty in neural networks, which we employ here. In our experiments, we rely on MC Dropout \cite{mcdropout} and ensembles \cite{deepensembles} to provide epistemic uncertainty estimates. As for the acquisition function, we use Upper Confidence Bound \cite{ucb} and Expected Improvement \cite{ei}. With the experiments of Section~\ref{sec:uncertaintyresults}, we study the effects of these choices and observe the improvement provided by incorporating the uncertainty estimates.

\section{Related Work}
Biological sequence design has been approached with a wide variety of methods: reinforcement learning \cite{angermueller2019model}, Bayesian optimization \citep{wilson2017reparameterization, belanger2019biological, moss2020bayesian, bayes4DD, bayesDD2}, search/sampling using deep generative models~\citep{brookes2019conditioning,kumar2019model,boitreaud2020optimol,das2021accelerated, hoffman2021optimizing,melnyk2021benchmarking}, deep model-based optimization~\citep{trabucco2021conservative}, adaptive evolutionary methods ~\citep{hansen2006cma, swersky20amortized, sinai2020adalead}, likelihood-free inference~\citep{Zhang2021UnifyingLI}, and black-box optimization with surrogate models ~\citep{dadkhahi2022fourier}.
As suggested in Section \ref{sec:gflownets}, GFlowNets have the potential to improve over such methods by amortizing the cost of search (e.g., when comparing with MCMC's mixing time) over learning, giving probability mass to the entire space facilitating exploration and diversity (vs e.g., RL which tends to be greedier), enabling the use of imperfect data (vs e.g., generative models that require strictly positive or negative samples), and by scaling well with data by exploiting structure in function approximation (vs e.g., Bayesian methods that can cost $\mathcal{O}(n^3)$ for  $n$ datapoints).

\section{Experiments}
\label{sec:experiments}
In this section we present experimental results across various biologically relevant sequence design tasks to demonstrate the effectiveness of our proposed GFlowNet-AL algorithm. We design our experiments to reflect realistic sequence design scenarios, varying several key parameters: 
\begin{enumerate}[leftmargin=*,itemsep=0pt, parsep=0pt,topsep=4pt]
    \item $N$: the number of active learning rounds - This can vary depending upon the particular application being considered, where the cost of evaluation in each round can limit the number of rounds available. 
    \item $b$: the size of candidate batch to be generated - The experimental setup in the evaluation phase can only be scaled to certain batch sizes, for instance, the synthesis of small molecules is mostly manual and cannot be parallelized much, whereas peptide synthesis can be scaled to $10^4$ to $10^6$  sequences at a time. 
    \item $|\mathcal{D}_0|$: the initial dataset available - Depending on the task at hand, one can have access to different numbers of initially available candidates. 
    \item $|x|$: the maximal length of constructed sequences - This can vary depending on the task at hand, for instance, design of anti-microbial peptides uses proteins of length $50$ or shorter, whereas design of fluorescent proteins uses sequences of length $>200$.
    \item $|\mathcal{A}|$: the size of the action space (vocabulary) - Depending on the type of biological sequence being considered the vocabulary size can vary, for instance, from $4$, in the case of DNA sequences, to $20$ in the case of proteins. 
\end{enumerate}

\subsection{Tasks and Evaluation Criteria}
\label{sec:task}
We present results on the following sequence design tasks. See Appendix~\ref{app:task} for further details on each of the tasks. 
\begin{itemize}[leftmargin=*,topsep=2mm,parsep=1mm]
    \item \textbf{Anti-Microbial Peptide Design}: The goal is to generate peptides (short protein sequences) with anti-microbial properties. We consider sequences of length $50$ or lower. The vocabulary size is $20$ (amino-acids). We consider $N=10$ rounds, with batch size $b=1000$ and starting dataset $\mathcal{D}_0$ with 3219 AMPs and 4611 non-AMP sequences from the DBAASP database~\citep{pirtskhalava2021dbaasp}. The choice of parameters was guided by the fact that AMPs can be efficiently synthesized and evaluated \emph{in-vitro} in large quantities. Details in Appendix~\ref{app:amptask}
    \item \textbf{TF Bind 8}: We follow \citet{trabucco2021designbench}, where the goal is to search the space of DNA sequences (vocabulary size $4$ nucleobases) of length $8$ that have high binding activity with human transcription factors. Following the offline Model-Based Optimization setting from \citet{trabucco2021designbench}, we consider a single round setting $N=1$, and generate $b=128$ candidates starting with $|\mathcal{D}_0| = 32,898$ examples. The data and oracle are from \cite{barrera2016survey}. Details in Appendix~\ref{app:tfbindtask}
    \item \textbf{GFP}: We use the design task as presented in \citet{trabucco2021designbench}. The goal is to search the space of protein sequences (vocabulary size $20$) of length $237$ and have high fluorescence. Following the offline Model-Based Optimization setting from \citet{trabucco2021designbench}, we consider a single round setting $N=1$, and generate $b=128$ candidates starting with $|\mathcal{D}_0| = 5,000$ examples. The data and oracle are from \cite{sarkisyan2016local, tape2019}. Details in Appendix~\ref{app:gfptask}. 
\end{itemize}

To evaluate the performance on these tasks, we follow the desiderata defined in Section~\ref{sec:problem}. We evaluate the Performance, Diversity and Novelty Scores on the highest-scoring generated candidates, $\dbest$. For the TF Bind 8 and GFP task we also present the 100th percentile and 50th percentile scores on the generated batch, following the evaluation scheme presented in \citet{trabucco2021designbench} in Appendix~\ref{app:additional_res}. 

\subsection{Baselines and Implementation}
\label{sec:baselines}
We consider as baselines a representative set of prior work focusing on ML for sequence design. We use the following methods as baselines: DynaPPO \citep[][Active Learning with RL as Generator]{angermueller2019model},  AmortizedBO \citep[][Bayesian Optimization with RL-based Genetic Algorithm for optimizing acquisiton function]{swersky20amortized}, and COMs \citep[][Deep Model Based Optimization]{trabucco2021conservative}. We also include a GFlowNet baseline with neither offline data nor uncertainty from the proxy. In all the experiments, the data is represented as a sequence of one-hot vectors $\{0, 1\}^{(|x|\times |\mathcal{A}|)}$, similar to the procedure followed by \citet{trabucco2021designbench}. For a fair comparison, we restrict all the baselines to use the same architecture (MLPs), however due to the large number of design choices in each of the baselines, there are some discrepancies. We provide complete implementation details in the Appendix~\ref{app:baselines}. 

\subsection{Results}
\label{sec:results}
\subsubsection{Anti-Microbial Peptide Design}
\label{sec:res_amp_main}

\begin{table}[]
\caption{Results on the AMP Task with $K=100$. }
\resizebox{\linewidth}{!}{
\begin{tabular}{lccc}
\hline
                      & \textbf{Performance} & \textbf{Diversity} & \textbf{Novelty} \\ \hline
\textbf{GFlowNet-AL}  & $\mathbf{ 0.932 \pm 0.002}$ & $\mathbf{ 22.34 \pm 1.24}$ & $\mathbf{ 28.44 \pm 1.32}$ \\
\textbf{DynaPPO}      & $\mathbf{0.938 \pm 0.009}$ & $12.12 \pm 1.71$ & $9.31 \pm 0.69$\\
\textbf{COMs}         & $0.761 \pm 0.009$ & $19.38 \pm 0.14$ & $26.47 \pm 1.3$ \\ 
\textbf{GFlowNet}      & $0.868 \pm 0.015$ & $11.32 \pm 0.67$ & $15.72 \pm 0.44$\\ \hline
\end{tabular}
}
\label{tab:amp_res100}
\end{table}
Table~\ref{tab:amp_res100} shows the the results for the AMP design task. 
We observe that GFlowNet-AL generates significantly more diverse and novel sequences compared to the baselines,
as well as better final TopK performance. Note that our
experiments with AmortizedBO here were not conclusive, 
as it was designed for fixed-length sequences and generated
nonsensical peptides (with almost exclusively W's and C's). 
See Appendix~\ref{app:amp_add} for examples of the sequences generated by AmortizedBO for this task. Another interesting observation here, in the setting of generating large batches, is that COMs, which relies on generating novel candidates by optimizing known candidates against a learned conservative model, performs quite poorly. This can be attributed the fact that it essentially performs a local search around known candidates, and this can be detrimental in cases where the goal is to generate large diverse and novel batches.

\begin{figure}
    \centering
    \includegraphics[width=0.8\linewidth]{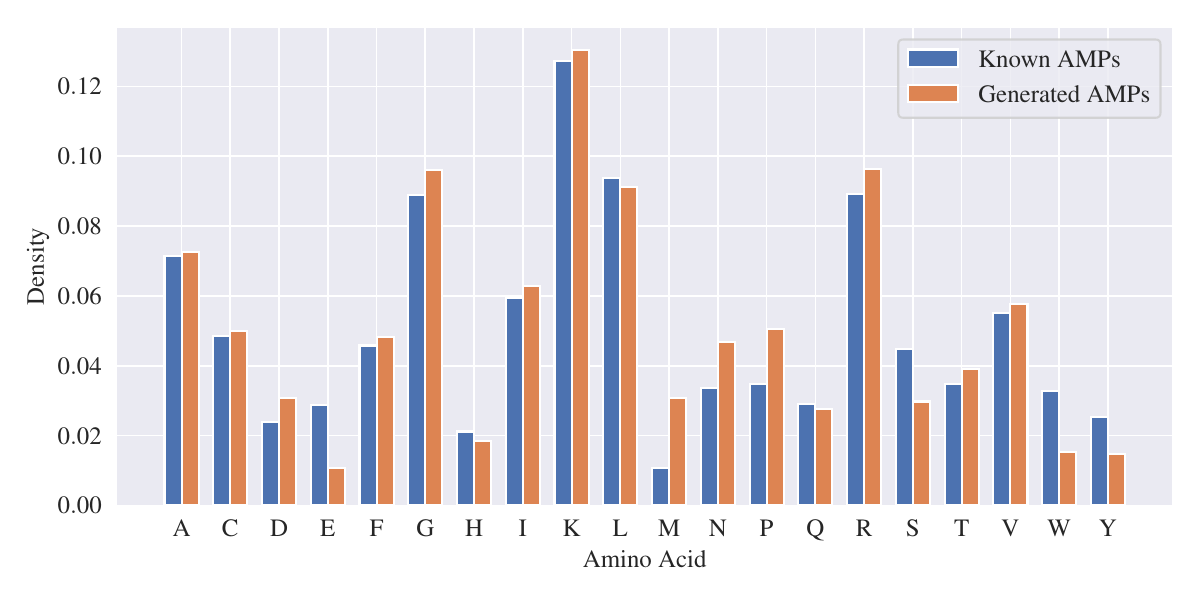}
    \caption{Distribution of occurrence of amino acids in the peptides generated with GFlowNet-AL closely matches that of known AMPs.}
    \label{fig:aa_dist}
\end{figure}

\paragraph{Physiochemical Properties}
In addition to the usefulness metric, to understand the biological relevance of the sequences generated by GFlowNet-AL we study several physiochemical properties of the Top 100 generated sequences using \texttt{BioPython} \cite{cock2003biopython}. The instability index for the generated peptides is 26.5 on averge with maximum of 36 (score of over 40 indicates instability). Figure~\ref{fig:aa_dist} shows the distribution of AAs in the generated sequences plotted against the set of known AMPs. We can observe that the distribution of amino acids in the generated sequences closely matches that of known AMPs. 

\begin{table}[h]
\caption{Results on TF-Bind-8 task with $K=128$}
\resizebox{\linewidth}{!}{
\begin{tabular}{llll}
\hline
                      & \textbf{Performance} & \textbf{Diversity} & \textbf{Novelty} \\ \hline
\textbf{GFlowNet-AL}  &  $\mathbf{0.84 \pm 0.05}$ & $4.53 \pm 0.46$ & $\mathbf{2.12 \pm 0.04}$\\
\textbf{DynaPPO}      &     $0.58 \pm 0.02$ & $5.18 \pm 0.04$  & $0.83 \pm 0.03$ \\
\textbf{COMs}         &  $0.74 \pm 0.04$ & $4.36 \pm 0.24$ & $1.16 \pm 0.11$ \\
\textbf{BO-qEI}         &  $0.44 \pm 0.05$ & $4.78 \pm 0.17$ & $0.62 \pm 0.23$ \\
\textbf{CbAS}         &  $ 0.45 \pm 0.14$ & $5.35 \pm 0.16$ & $0.46 \pm 0.04$ \\
\textbf{MINs}         &  $0.40 \pm 0.14$ & $\mathbf{ 5.57 \pm 0.15}$ & $0.36 \pm 0.00$ \\
\textbf{CMA-ES}         &  $0.47 \pm 0.12$ & $4.89 \pm 0.01$ & $0.64 \pm 0.21$ \\
\textbf{AmortizedBO} &   $0.62 \pm 0.01$               &     $4.97 \pm 0.06$               &         $1.00 \pm 0.57$         \\ 
\textbf{GFlowNet}      & $0.72 \pm 0.03$ & $4.72 \pm 0.13$ & $1.14 \pm 0.3$\\ \hline
\end{tabular}}
\label{tab:tfbind}
\end{table}

\subsubsection{TF-Bind-8}
The TF-Bind-8 task requires searching in the space of short DNA sequences for high binding activity with human transcription factors. The initial dataset $\mathcal{D}_0$ consists of the lower scoring half of all the possible DNA sequences of length $8$. This setup allows us to evaluate the methods in the common setting, where only low quality data is available initially. For this task, we also include additional MBO baselines presented in \citet{trabucco2021designbench}. 
On this task, we see from Table~\ref{tab:tfbind} that GFlowNet-AL performs better than the other baselines in terms of TopK performance and novelty but that the MINs method performed best in terms of diversity. However, when we look at all the metrics together, MINs have a much lower performance score and novelty score indicating they generate sequences mostly from the training set. We also present results on the 100th and 50th percentile metrics proposed in \citet{trabucco2021conservative}, in the Appendix~\ref{app:tfb_gfp_add}, where GFlowNet still outperforms all the evaluated methods.

\subsubsection{GFP}
Finally we consider the GFP task, where the goal is to search in the space of proteins with for proteins that are highly fluorescent. Similar to TF-Bind-8, we include baselines from \citet{trabucco2021conservative}. In this task we observe that GFlowNet-AL does not outperform the baselines\footnote{Upon publication, several issues were identified with the results on this task, including sensitivity to initial conditions for the proxy and an error in the computation of the diversity and novelty. The results here have been updated to reflect these issues.}. The data used in the task consists of mutations of a single protein. When the proxy is trained on a biased subset of this data, it can have spurious modes which are not present in the true reward. As the proxy defines the reward optimized by the generative policy, de novo generation with GFlowNet-AL and DynaPPO struggles to find good candidates. 

\begin{table}[h]
\caption{Results on GFP task with $K=128$}
\resizebox{\linewidth}{!}{
\begin{tabular}{llll}
\hline
                      & \textbf{Performance} & \textbf{Diversity} & \textbf{Novelty} \\ \hline
\textbf{GFlowNet-AL}  & $0.05 \pm 0.010$ & $21.57 \pm 3.73$ & $ 31.52 \pm 2.82$ \\
\textbf{DynaPPO}      & $0.05 \pm 0.008$ & $12.54 \pm 1.34$ & $15.10 \pm 3.37$\\
\textbf{COMs}         & $\mathbf{0.831\pm 0.003}$ & $8.57 \pm 1.21$ & $10.31 \pm 1.45$\\
\textbf{BO-qEI}         & $0.045 \pm 0.003$  & $12.87\pm 1.09$ & $22.88 \pm 4.62$  \\
\textbf{CbAS}         & $0.817 \pm 0.012$  & $8.53 \pm 0.65$ & $8.72 \pm 1.26$  \\
\textbf{MINs}         & $0.761 \pm 0.007$ & $8.31 \pm 0.02$  & $4.45 \pm 0.52$ \\
\textbf{CMA-ES}         & $0.063 \pm 0.003$  & $10.52 \pm 4.24$ & $10.77 \pm 4.12$ \\
\textbf{AmortizedBO} & $0.051 \pm 0.001$ &  $16.14 \pm 2.14$ & $19.31 \pm 2.43$ \\
\end{tabular}}
\label{tab:gfp}
\end{table}

\subsection{Ablations}
\subsubsection{Training with the oracle data}
\label{sec:mixingresults}
\begin{figure}
    \centering
    \includegraphics[width=\linewidth]{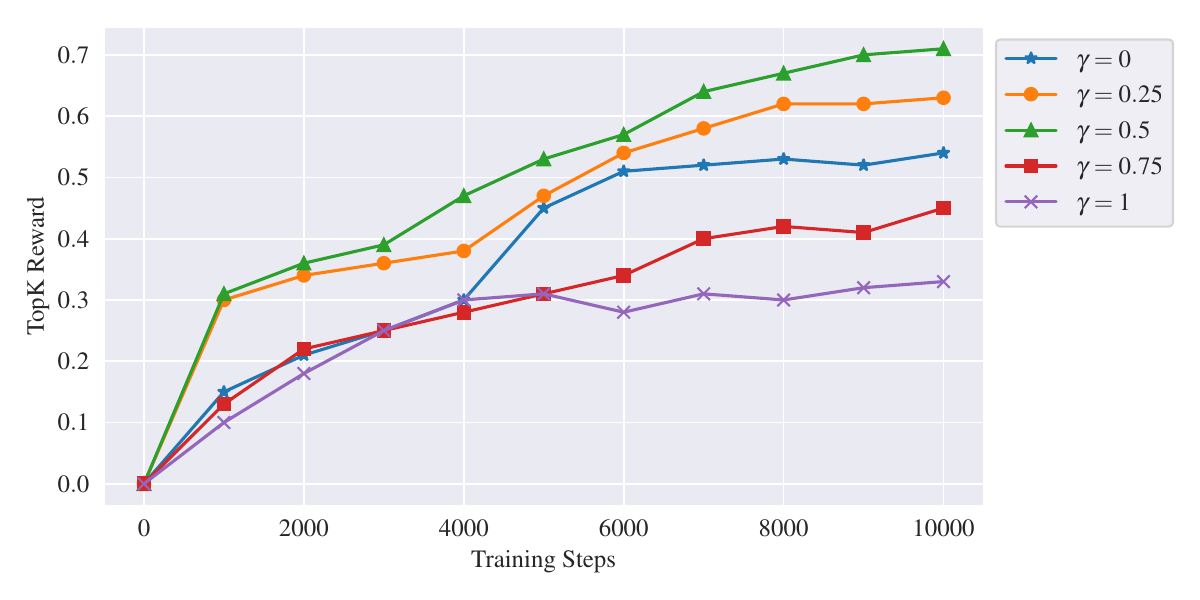}
    \caption{TopK ($K=100$) scores over the training iterations for GFlowNet-AL in Round 1 of AMP Generation Task, with different values of $\gamma$, the proportion of trajectories sampled from the data. We can observe that $\gamma=50\%$ is best.}
    \label{fig:empdist_effect}
\end{figure}

We perform ablations to isolate the effect of including trajectories sampled from a static dataset in the GFlowNet training procedure, discussed in Algorithm~\ref{algo:inner_loop}. To do this, we sample a set of $4096$ examples every $1000$ training steps for the GFlowNet, and consider the average reward of the Top100 sequences in that set. Figure~\ref{fig:empdist_effect} shows the progression of the Top100 scores over the course of training of the GFlowNet in the first round of active learning in the AMP task, for different values of $\gamma$, which represents the fraction of sequences sampled from the dataset within a mini-batch. As we increase $\gamma$ away from 0, the performance improves significantly compared to not having any offline data ($\gamma=0$), until $\gamma=0.50$ which worked best. Too many dataset examples, i.e., too few on-policy trajectories, leads to less exploration and less generalization outside of the training examples. Using offline data improves the speed at which GFlowNet training covers the support of the optimal $\pi$. Going beyond $\gamma=0.5$, however, performance becomes significantly worse. 

\subsubsection{Effect of uncertainty estimates}
\label{sec:uncertaintyresults}
Next, we study the effect of incorporating information about the uncertainty in the {\em learned} reward function through multiple rounds of active learning. Here again, we consider the AMP Generation task, and consider three variations of the GFlowNet-AL algorithm with different models $M$ in the proxy. We consider Deep Ensembles \cite{deepensembles} and MC Dropout \cite{mcdropout} as two representative uncertainty estimation methods for neural networks and a third variation with a single model for the proxy, corresponding to the case of not having the uncertainty of the model incorporated in the reward. Note that Deep Ensembles generally provide more accurate uncertainty estimates than MC Dropout, so our evaluation also covers the effect of the quality of the uncertainty estimates. Table~\ref{tab:uncertaintyeffect} shows the results for these ablations. We can observe that having any uncertainty estimate can provide an advantage over having none. In addition, we also observe that more accurate uncertainty estimates from Deep Ensembles lead to better results overall. We present additional ablations on the effect of the acquisition function in the Appendix~\ref{app:additional_res}, but note that we do not see a significant difference based on the choice of the acquisition function.

\begin{table}[H]
\centering
\caption{Results on the AMP Task with $K=100$ for GFlowNet-AL with different methods for uncertainty estimation, with UCB as the acquisition function. }
\resizebox{\linewidth}{!}{
\begin{tabular}{lccc}
\hline
                      & \textbf{Performance} & \textbf{Diversity} & \textbf{Novelty} \\ \hline
\textbf{GFlowNet-AL Ensemble}  & $\mathbf{ 0.932 \pm 0.002}$ & $\mathbf{ 22.34 \pm 1.24}$ & $\mathbf{ 28.44 \pm 1.32}$ \\
\textbf{GFlowNet-AL MC Dropout}      & $0.921 \pm 0.004$ & $18.58 \pm 1.78$ & $19.58 \pm 1.12$\\
\textbf{GFlowNet-AL None}      & $0.909 \pm 0.008$ & $16.42 \pm 0.74$ & $17.24 \pm 1.44$\\
\hline
\end{tabular}
}
\label{tab:uncertaintyeffect}
\end{table}

\section{Conclusion and Future Work}
Motivated by global health challenges such as antimicrobial resistance and the currently expensive and slow process of discovering new and useful biological sequences, we have introduced a generative active learning algorithm for sequences based on GFlowNets (as the candidate generator) and principles from Bayesian optimization (the estimation of epistemic uncertainty and the use of an acquisition function to score candidates), with the objective to produce diverse and novel sets of candidates. To achieve this, we discovered training GFlowNets could be greatly accelerated by using training sequences from the oracle (e.g., biological experiments) to construct additional training trajectories. We validated that both the use of epistemic uncertainty and the empirical distribution derived from the oracle outputs helped to obtain better results, especially in terms of diversity and relative novelty of the generated candidates.
A \textbf{limitation} of the proposed method and others that involve both a proxy model and a generative policy is that we now have two separate learners, each with their hyper-parameters. On the other hand, the use of efficient optimization or generation is necessary in high-dimensional search spaces. We also note the poor performance on the GFP task with de novo generation struggling to find good candidates.
\textbf{Future work} should explore how we can make the retraining of the proxy model more efficient, considering that this is a continual learning setting. Better estimators of information gain, non-autoregressive generative models taking advantage of the underlying structure in the data, and an outer loop policy handling multiple oracles with a different fidelity are natural extensions of this work. 

{\bf Software and Data:} The code is available at \href{https://github.com/MJ10/BioSeq-GFN-AL}{https://github.com/MJ10/BioSeq-GFN-AL}.



\section*{Acknowledgements}
The authors would like to thank Dianbo Liu, Xu Ji, Kolya Malkin, Leo Feng, and members of the Drug Discovery and GFlowNet groups at Mila as well as anonymous reviewers for helpful discussions and feedback. The authors also acknowledge support from the AIHN IBM-MILA project. This research was enabled in part by compute resources provided by Compute Canada. The authors acknowledge funding from CIFAR, Samsung, IBM, Microsoft. 

\bibliography{main}
\bibliographystyle{icml2022}

\newpage
\appendix
\onecolumn
\section{Task Details}
\label{app:task}
\subsection{Anti-Microbial Peptides}
\label{app:amptask}
The peptides used in our experiments are obtained by filtering DBAASP \citep{pirtskhalava2021dbaasp}. We select peptides with sequence length between 12 and 60 as well as choosing unusual amino acid to the type of ``without modification''.  The target group is the Gram-positive bacteria. In total we have 6438 positive AMPs, and 9522 non-AMPs.

We split the above mentioned dataset into two parts: $D_1$ and $D_2$. $D_1$ is available for use the algorithms, whereas, $D_2$ is used to train the oracle, $f$, following \citep{angermueller2019model}, as a simulation of wet-lab experiments for the generated sequences. 
Notice that every observation in the dataset has its corresponding group. The definition of being in the same group could be: having the same target or the same title or the same cluster. We follow a strict principle to split the dataset into $D_1$ and $D_2$: for any observation $x$ in $D_1$, there are no observations in $D_2$ belong to $x$'s group, and vice versa. Under this principle, the $D_1$ and $D_2$ are made either by cross-validation split or by train-valid split.
Unlike \citep{angermueller2019model}, we use MLP classifiers (up to 89\% test accuracy) to train the oracles based on features with the pre-trained protein language models from \cite{elnaggar2020prottrans}, instead of Random Forests. Because the lengths of the sequences are not fixed, we set 60 as the maximum length of the sequences. We pad the sequences which do not reach the length of 60 by appending the end of sequence token.

\subsection{TF-Bind-8}
The dataset used for the TF-Bind-8 task contains 65792 samples, representing every possible size 8 string of nucleotides $\boldsymbol{x} \in \{0, 1\}^{8 \times 4}$. 50\% of the initial dataset is set aside for model training, resulting in a training set of size 32898. As the dataset includes all possible size 8 DNA sequences, the oracle for this task is exact. The dataset for the TF-Bind-8 task is derived from~\cite{barrera} wherein  DNA sequences are scored based on their binding activity to a human transcription factor SIX6 REF R1, where a higher binding energy is better. This task has been used to demonstrate MBO algorithm performance in recent papers~\cite{angermueller2019model,trabucco2021conservative}.  We use the implementation of this dataset from the Design-Bench repository without any preprocessing (\url{https://github.com/brandontrabucco/design-bench}).  The dataset's construction is described in more detail in~\cite{trabucco2021conservative}.   
\label{app:tfbindtask}

\subsection{GFP}
\label{app:gfptask}
We again use the implementation in the Design-Bench repository for the GFP task's dataset and oracle (\url{https://github.com/brandontrabucco/design-bench})~\cite{trabucco2021designbench}.  The GFP task requires generation of proteins derivative of the bio-luminescent jellyfish \textit{Aequorea victoria}'s green fluorescent protein (GFP) with maximum fluorescence.  The dataset is of size 56086 with each sample being a protein of length 237. Each protein is encoded as a tensor of 237 sequential one-hot vectors, written as $\boldsymbol{x} \in \{0,1\}^{237 \times 20}$.  While the full dataset is of size 56086 only 5000 samples, drawn from between the 50th and 60th percentiles, are given as a training set to the optimization algorithms. The oracle used is 12-layer transformer provided by the TAPE framework~\cite{Rao2019EvaluatingPT}. Before running any optimization algorithm, we normalize the fluorescence scores given in the training set and produced by the oracle. Finally, in reporting final scores we re-normalize with respect to the full GFP dataset's minimum and maximum.  More details on the setup are provided in~\citet{trabucco2021conservative}.
\section{Implementation Details}
\subsection{Baselines}
\label{app:baselines}
In our implementations of the baseline algorithms, we made use of previously published implementations, making small adaptations where necessary. In particular, for AmmortizedBO we used their published implementation and for DynaPPO we adapted and used a version implemented in the repository published by the FLEXS project~\cite{sinai2020adalead}.

\begin{table}[H]
\centering
\caption{Hyperparameters used for AmortizedBO.  We varied the number of mutations allowed, $K$, based on the length of the sequence to be generated, $L$.  We also varied the acquisition function used, as well as the number of generations $G$ allowed between proposals.}
\begin{tabular}{lcccc}
\hline
                 &   \textbf{$L$}  & \textbf{$K$} & \textbf{Acquisition Function} & \textbf{$G$} \\ \hline
\textbf{AMP}  & $50$ & $40$ & UCB & $40$ \\
\textbf{TF-Bind-8}      & $8$ & $4$ & UCB & $5$\\
\textbf{GFP}         &   $237$                   &   $200$                 &  UCB & $5$                \\ \hline
\end{tabular}
\label{tab:am_bo_hparams}
\end{table}

\textbf{AmortizedBO:} Following~\cite{swersky20amortized} we kept nearly all hyperparameters constant across all tasks for AmmortizedBO, only varying the hyperparameters listed in Table~\ref{tab:am_bo_hparams}. Those hyperparameters were selected after a grid search for which the same options were provided for each task.  For all other hyperparameters and architectures we use the default settings in the published AmortizedBO implementation.  For the AMP task we required that AmortizedBO may output dynamically sized strings.  To implement this, we allowed AmortizedBO to output a stop token at any position which would cause that position to the maximum length of the string to be padded.  This setting was only used for the AMP task.

\begin{table}[H]
\centering
\caption{Hyperparameters used for DynaPPO.  We varied the number of trajectories generated between proposals $\Tau$, the policy network learning rate $\gamma$, the DynaPPO exploration penalty scale factor $\lambda$, and the exploration penalty's radius $\epsilon$.}
\begin{tabular}{lcccc}
\hline
                 &   \textbf{$\Tau$}  & \textbf{$\gamma$} & \textbf{$\lambda$} & \textbf{$\epsilon$} \\ \hline
\textbf{AMP}  & $1000$ & $0.0001$ & $0.2$ & $8$ \\
\textbf{TF-Bind-8}      & $20000$ & $0.0001$ & $0.1$ & $2$\\
\textbf{GFP}         &   $2000$                   &   $0.0001$                 &  $0.1$ & $20$                \\ \hline
\end{tabular}
\label{tab:dynappo_hparams}
\end{table}

\textbf{DynaPPO:} Although we used the FLEXS library as our base for DynaPPO, we altered the implementation slightly in order to better represent the algorithm specified in~\cite{angermueller2019model}.  In particular, we added dynamic hyperparameter tuning of the proxy model after each query to the oracle, the exploration penalty term proposed by DynaPPO, and a method to allow DynaPPO to output variably sized strings (for the AMP task). We reused the architecture specified in the FLEXS library for the policy network and, as DynaPPO requires a hyperparameter search across its proxy model after each query to the oracle, the hyperparameter options for the proxy laid out in~\cite{angermueller2019model} in our implementation.  We ran a grid search over various settings of the number of trajectories generated between proposals $\Tau$, the policy network learning rate $\gamma$, the exploration penalty scale parameter $\lambda$, and the exploration penalty radius $\epsilon$.  We used the best performing hyperparameters for each task, as reported in Table~\ref{tab:dynappo_hparams}.  In DynaPPO's proxy ensemble, a constituent model is only included in the ensemble if its $R^2$ score is higher than a threshold given $5$-fold cross validation on the dataset.  Following their recommendation, we require a model's $R^2$ on the dataset to be at least $0.5$ for it to be included in the ensemble. Finally, we note that \citet{angermueller2019model} propose including a Gaussian Process in their proxy ensemble.  However, we found the Gaussian Process to be intractable on the datasets for our experimental tasks, and as such excluded it from the proxy ensemble.

\textbf{COMs}:
For COMs we the same architecture for the forward model as used in \citet{trabucco2021conservative}, which is an feedforward neural network with two hidden layers of size $2048$ and a LeakyRELU activation function with leak 0.3, trained with an Adam Optimizer and learning rate $10^-3$. The rest of the parameters are set to the best hyperparameters reported in \citet{trabucco2021conservative} for all tasks as follows: number of gradient ascent steps in the solver=50, number of steps to generate adversarial $\mu(x)$ = 50, learning rate $\alpha = 0.01$, $\tau = 2.0$, $\eta = 2\sqrt{d}$ and number of epochs to train $\hat{f}_\theta$ = 50, as well as the same pre-processing procedure.  

\textbf{Other Baselines}: 
For the additional baselines reported on the GFP and TF-Bind-8 tasks: MINs \cite{kumar2019model}, CbAS \cite{Brookes2019ConditioningBA}, BO-qEI \cite{wilson2017reparameterization} and CMA-ES \cite{hansen2006cma}, we use the the implementations provided in \citet{trabucco2021designbench}, and reproduce the results with the reported hyperparameters.

\subsection{GFlowNet}
\label{app:gfn_impl}
We implement the proposed GFlowNet-AL algorithm in PyTorch \cite{paszke2019pytorch}. 

\textbf{Proxy}: 
We use MLP with 2 hidden layers of dimension $2048$ and ReLU activation, as the base architecture for the proxy in our experiments in all three tasks. In the case of ensembles, we use 5 members with the same architecture, whereas in the case of MC Dropout we use $25$ samples with dropout rate $0.1$, and weight decay of $0.0001$. We use a minibatch size of $256$ for training with a MSE loss, using the Adam optimizer \cite{kingma2017adam}, with learning rate $10^-4$ and $(\beta_0, \beta_1)=(0.9, 0.999)$. We use early stopping, keeping $10\%$ of the data as a validation set. For UCB ($\mu + \kappa \sigma$) we use $\kappa=0.1$. 

\textbf{GFlowNet Generator}:
We parameterize the flow as a MLP with 2 hidden layers of dimension $2048$, and $\mathcal{A}$ outputs corresponding to each action. We use the trajectory balance objective for training in all our experiments. For training we use the Adam optimizer with $(\beta_0, \beta_1)=(0.9, 0.999)$. Table~\ref{tab:gfn_params} shows the rest of the hyperparameters. In addition to that we set $\gamma$, the proportion of offline trajectories to $0.5$ for all three tasks. The learning rate for $\log Z$ is set to $10^{-3}$ for all the experiments. In each round we sample $t * K$ candidates, and pick the top $K$ based on the proxy score, where $t$ is set to $5$ for all experiments.

\begin{table}[H]
\centering
\caption{Hyperparameters for the GFlowNet Generator}
\begin{tabular}{llll}
\hline
\textbf{Hyperparameter}              & \textbf{AMP} & \textbf{TF-Bind-8} & \textbf{GFP} \\ \hline
$\delta$: Uniform Policy Coefficient & 0.001        & 0.001              & 0.05         \\
Learning rate                        & $5 \times 10^{-4}$    & $10^{-5}$              & $5\times 10^{-4}$       \\
$m$: Minibatch size                  & 32           & 32                 & 32           \\
$\beta$: Reward Exponent $R(x)^{\beta}$              & 3            & 3                  & 3            \\
$T$: Training steps  & 10,000       & 5,000              & 20,000       \\ \hline
\end{tabular}
\label{tab:gfn_params}
\end{table}

For the results in Table~\ref{tab:amp_res100}, Table~\ref{tab:tfbind} and Table~\ref{tab:gfp}, we use GFlowNet-AL with ensembles as the proxy model with UCB as the acquisiton function. 

\section{Additional Results}
\label{app:additional_res}
\subsection{AMP Generation: Additional Results}
\label{app:amp_add}

As discussed in Section~\ref{sec:res_amp_main}, after running AmortizedBO on the AMP task the algorithm generated sequences which were overwhelmingly poor in regards to real-world usefulness. As AmortizedBO was initially proposed as an algorithm to produce fixed length strings, we implemented a dynamic length AmortizedBO for which we added an extra stop token to the generator's vocabulary.  When AmortizedBO selected to insert a stop token at position $i$,the selected position as well as all positions succeeding the selected position would be set to a padding token. 
All top 1000 sequences generated by AmortizedBO were of the maximum allowed sequence length for the AMP task. Some sequences are:
"RRRRWWRHHHHHICCWIWKCWWWIIICWWWWWCWWWWWIIWWWIIWCWWL", 
"RRRWRWWHHHWIICCHCIKCCLWIIIIWWWWWWCWWWWIWWWWIICWWWL", and 
"RRRWRWICHHRRCCCRIIWCCLWIIICWWWWWCWWWWIIWWWWIWCWWWL". These sequences do not look natural, and lack important characteristics generally found in AMPs (for example the amino acid "K", which is dominant in peptides with anti-microbial activity).

\subsection{TF-Bind-8 and GFP: Additional Results}
\label{app:tfb_gfp_add}
In Table~\ref{tab:max_med_gfp} and Table~\ref{tab:max_med_tf_bind} we present the 100th and 50th percentile results on the GFP and TF-Bind-8 tasks respectively as proposed in \citet{trabucco2021designbench}. We observe that GFlowNets outperform the baselines even under these metrics. 

\begin{table}[H]
\centering
\caption{Maximum and median scores of the proposed sequences for the GFP task.}
\begin{tabular}{lll}
\hline
                     & \textbf{100th Percentile} & \textbf{50th Percentile} \\ \hline
\textbf{GFlowNet-AL} &  $\mathbf{ 0.871\pm 0.006}$ & $\mathbf{ 0.853 \pm 0.002}$ \\
\textbf{DynaPPO}     &  $0.790\pm 0.003$ & $0.790 \pm 0.005$\\
\textbf{COMs}        &        $0.864 \pm 0.000$                   &     $0.864 \pm 0.000$                     \\
\textbf{BO-qEI}      &          $0.254 \pm 0.352$                 &     $0.246 \pm 0.341$                     \\
\textbf{CbAS}        &        $0.865 \pm 0.000$                   &     $0.852 \pm 0.004$                     \\
\textbf{MINs}        &          $0.865 \pm 0.001$                 &     $0.820 \pm 0.018$                     \\
\textbf{CMA-ES}      &       $0.054 \pm 0.002$                    &      $0.047 \pm 0.000$                    \\
\textbf{AmortizedBO} &   $0.058 \pm 0.002$     & $0.052 \pm 0.001$ \\ \hline
\end{tabular}
\label{tab:max_med_gfp}
\end{table}

\begin{table}[H]
\centering
\caption{Maximum and median score of the proposed sequences for the TF-Bind-8 task.}
\begin{tabular}{lll}
\hline
                     & \textbf{100th Percentile} & \textbf{50th Percentile} \\ \hline
\textbf{GFlowNet-AL} &  $\mathbf{0.989 \pm 0.009}$ &  $\mathbf{ 0.784 \pm 0.015}$\\
\textbf{DynaPPO}     &    $0.942 \pm 0.025$    &   $0.562 \pm 0.025$                   \\
\textbf{COMs}        &    $0.945 \pm 0.033$                       &      $0.497 \pm 0.038$                    \\
\textbf{BO-qEI}      &      $0.798 \pm 0.083$                     &          $0.439 \pm 0.000$                \\
\textbf{CbAS}        &        $0.927 \pm 0.051$                   &          $0.428 \pm 0.010$                \\
\textbf{MINs}        &          $0.905 \pm 0.052$                 &         $0.421 \pm 0.015$                 \\
\textbf{CMA-ES}      &          $0.953 \pm 0.022$                 &          $0.537 \pm 0.014$                \\
\textbf{AmortizedBO} &               $0.989 \pm 0.014$                   &   $0.636 \pm 0.025$                         \\ \hline
\end{tabular}
\label{tab:max_med_tf_bind}
\end{table}

\subsection{Effect of Uncertainty: Additional Results}
We present additional results on the three tasks with different choices of acquisition functions and uncertainty estimation methods. Note that for GFlowNet-AL-None, the choice of acquisition function does not matter, so the results are put only on the UCB section. We observe that the key factor affecting performance is consistently the uncertainty model.

\begin{table}[h!]
\centering
\caption{Results with GFlowNet-AL-None, where the uncertainty from the proxy is not used. }
\begin{tabular}{llll}
\hline
                   & \textbf{Performance} & \textbf{Diversity} & \textbf{Novelty} \\ \hline
\textbf{AMP}       & $0.909 \pm 0.008$ & $16.42 \pm 0.74$ & $17.24 \pm 1.44$ \\
\textbf{TF-Bind-8} & $0.81\pm 0.04$ & $3.96 \pm 0.32$ & $1.73 \pm 0.18$ \\
\textbf{GFP}       & $0.786\pm 0.001$ & $205.28 \pm 1.68$ & $207.65 \pm 1.19$ \\ \hline
\end{tabular}
\label{tab:gfn-al-none}
\end{table}

\begin{table}[h]
\centering
\caption{Results on AMP Generation Task with UCB and EI as acquisition functions and different methods for uncertainty estimation. }
\resizebox{\linewidth}{!}{
\begin{tabular}{lcccccc}
\hline
                               & \multicolumn{3}{c}{\textbf{UCB}}                             & \multicolumn{3}{c}{\textbf{EI}}                              \\ \cline{2-7} 
                               & \textbf{Performance} & \textbf{Diversity} & \textbf{Novelty} & \textbf{Performance} & \textbf{Diversity} & \textbf{Novelty} \\ \hline
\textbf{GFlowNet-AL-Ensemble}  & $\mathbf{0.932\pm 0.002}$ & $22.34 \pm 1.24$ & $\mathbf{ 28.44 \pm 1.32}$ & $0.928 \pm 0.002$ & $\mathbf{ 23.61 \pm 1.05}$ & $26.52 \pm 1.56$ \\
\textbf{GFlowNet-AL-MCDropout} & $0.921 \pm 0.004$ & $18.58 \pm 1.78$ & $19.58 \pm 1.12$ & $0.917 \pm 0.002$ & $17.38 \pm 0.64$ & $18.34 \pm 1.42$ \\

\end{tabular}
}
\end{table}

\begin{table}[h]
\centering
\caption{Results on TF-Bind-8 Task with UCB and EI as acquisition functions and different methods for uncertainty estimation. }
\resizebox{\linewidth}{!}{
\begin{tabular}{lcccccc}
\hline
                               & \multicolumn{3}{c}{\textbf{UCB}}                             & \multicolumn{3}{c}{\textbf{EI}}                              \\ \cline{2-7} 
                               & \textbf{Performance} & \textbf{Diversity} & \textbf{Novelty} & \textbf{Performance} & \textbf{Diversity} & \textbf{Novelty} \\ \hline
\textbf{GFlowNet-AL-Ensemble}  & $\mathbf{0.84 \pm 0.05}$ & $\mathbf{4.53 \pm 0.46}$ & $\mathbf{2.12 \pm 0.04}$ & $0.84 \pm 0.01$ & $4.46 \pm 0.58$ & $2.02 \pm 0.13$  \\
\textbf{GFlowNet-AL-MCDropout} & $0.81 \pm 0.03$ & $3.89 \pm 0.85$ & $1.76 \pm 0.15$ & $0.81 \pm 0.02$ & $4.10 \pm 0.43$ & $1.92 \pm 0.16$\\
\end{tabular}
}
\end{table}

\begin{table}[h!]
\centering
\caption{Results on GFP Task with UCB and EI as acquisition functions and different methods for uncertainty estimation.}
\resizebox{\linewidth}{!}{
\begin{tabular}{lcccccc}
\hline
                               & \multicolumn{3}{c}{\textbf{UCB}}                             & \multicolumn{3}{c}{\textbf{EI}}                              \\ \cline{2-7} 
                               & \textbf{Performance} & \textbf{Diversity} & \textbf{Novelty} & \textbf{Performance} & \textbf{Diversity} & \textbf{Novelty} \\ \hline
\textbf{GFlowNet-AL-Ensemble}  & $\mathbf{ 0.853\pm 0.004}$ & $211.51\pm 0.73$ & $\mathbf{ 210.56\pm 0.82}$ & $0.851 \pm 0.003$ & $\mathbf{ 212.03 \pm 0.64}$ & $208.31 \pm 0.94$  \\
\textbf{GFlowNet-AL-MCDropout} & $0.825 \pm 0.007$ & $204.76 \pm 1.75$ & $200.93 \pm 0.46$ & $0.838 \pm 0.001$ & $207.42 \pm 1.24$ & $208.31 \pm 1.60$ \\
\end{tabular}
}
\end{table}


\end{document}